\documentstyle[12pt,psfig]{article}  
\begin{document}
\date{}                  
\bibliographystyle{unsrt}
\def\br{{\bf r}}
\def\DF{\Delta F}
\def\bk{{\bf k}}
\def\bR{{\bf R}}
\def\<{\langle}
\def\>{\rangle}

\title{Macromolecular separation through a porous surface.}
\author{J.M.Deutsch\\ 
Physics Department, University of California, Santa Cruz CA 95064\\
and\\
Hyoungsoo Yoon\\
Basic Science Research Institute, Postech, Pohang, Korea 790-784.}
\maketitle
\abstract
{A new technique for the separation of macromolecules
is proposed and investigated. A thin mesh with pores
comparable to the radius of gyration of a free chain
is used to filter chains according to their length.
Without a field it has previously been shown that the
permeability decays as a power law with chain length.
However by applying particular configurations
of pulsed fields, it is possible to have a permeability
that decays as an exponential. This faster decay
gives much higher resolution of separation. We also
propose a modified screen containing an array of holes
with barb-like protrusions running parallel to the
surface. When static friction is present between the
macromolecule and the protrusion, some of the chains
get trapped for long durations of time. By using
this and a periodic modulation of an applied electric
field, high resolution can be attained.
}

\vskip 0.3 truein

\section{Introduction}

Recent advances in gel electrophoresis have dramatically
improved the separation of DNA molecules. The use of
pulsed electric fields has now made it possible to do
routine separation of mega-base DNA molecules. 
There are still improvements in the technique that would
be highly desirable. Here a variant of electrophoresis
is considered. Instead of separation through a gel,
we consider the separation of molecules through porous
surfaces, that look like fine screens. 

The obvious advantage of such an approach is that separation
does not require migration over distances long compared
to chain length, 
as is necessary with gel electrophoresis. In that case
typically a molecule drifts of order ten centimeters.
Instead, separation can only occur at the screen itself.
Thus effects of trapping that have caused problems
might be avoided. One might also hope that separation
utilizing a screen may be more rapid, as the total displacement
of the molecule is much smaller.  

Here we investigate two  ways  to separate macromolecules
at a surface. In many respects the findings are encouraging,
however there are still considerable technical difficulties
that will be necessary to overcome in order to make such
ideas practical. The most obvious difficulty is in actually
manufacturing such a porous screen. For the separation
of DNA, it would need to contain an array of holes of
size of order a few thousand angstroms. Its thickness
should also be of this order of magnitude. 

The motion of a polymer through such a screen has been
investigated previously\cite{yoon,obukhov}. It was found that
the permeability of the molecule decreases as $M^{-2}$
where $M$ is the molecular weight. It would be highly
desirable to have a more rapid decrease, as this would 
increase resolution.

In this paper, first
we present data for charged molecules moving near
a pore in the presence  of a pulsed electric field. 
With an appropriate
configuration of pulses, we find the permeability of
the device decreased very rapidly with chain length,
as an exponential for long enough chains. This then gives
rise to a sharp separation, one that is tunable, by 
varying the time-scale of pulses in the external field.

Second we present the outline of another device with
even sharper resolution. It involves both holes and
pegs, and may be impossible to fabricate with current
technology. However because of the rapid development
of fabrication techniques it may be practical in only
a few years. Such a device might then have interesting
applications.

\section{Surfaces with holes}
\label{sec:holes}

When a polyelectrolyte is close to a screen, it can
pass through it by a process similar to reptation\cite{degennes}.
If a weak field is applied perpendicular to the
surface, this
will lead to some differential separation of chains
according to their chain lengths. However the effect
is a power law\cite{yoon,obukhov}, and so is rather weak. Here we propose 
a method for increasing the power of separation dramatically. 

\begin{figure}[tbh]
\begin{center}
\                
\psfig{file=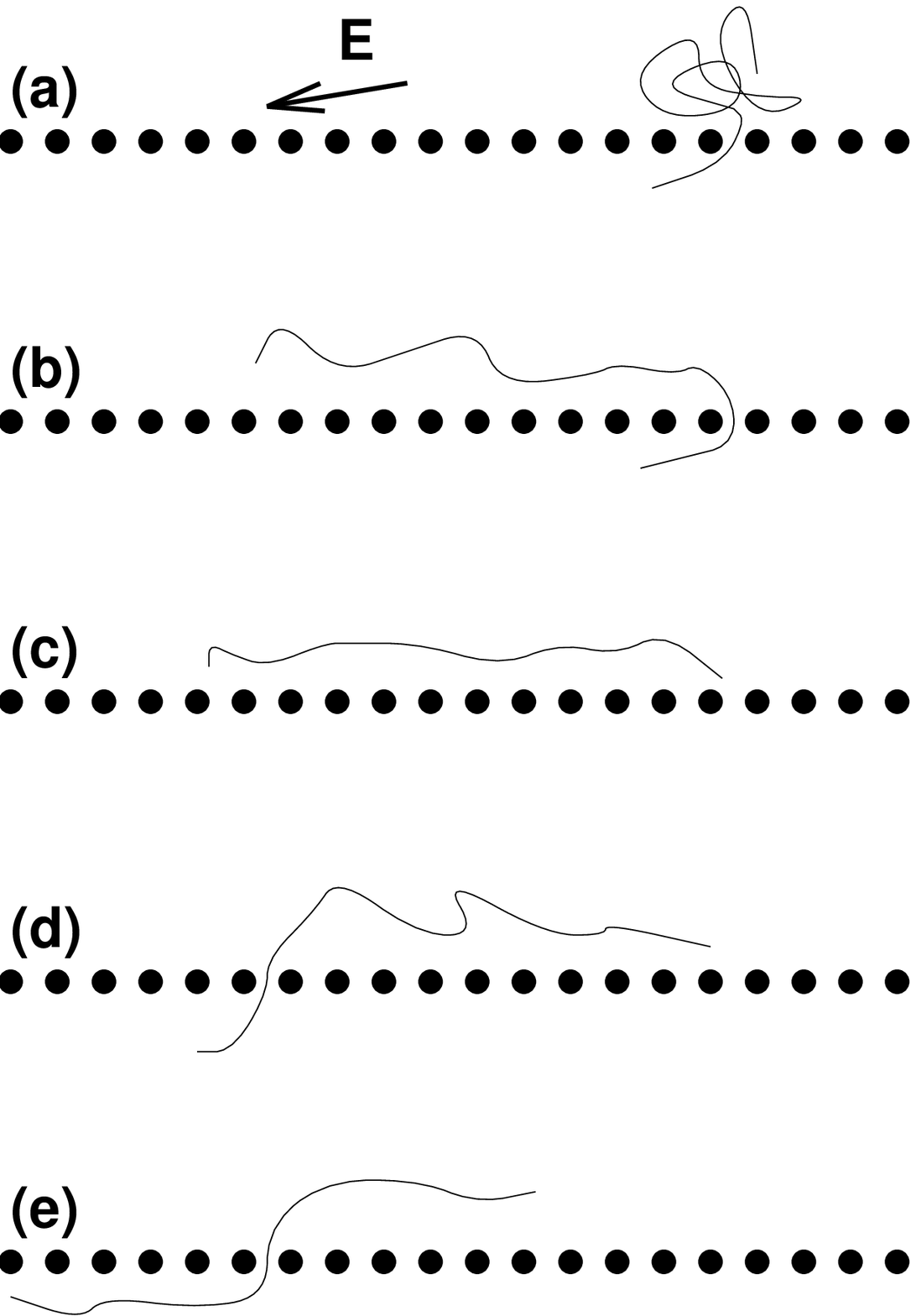,width=3in}
\end{center}
\caption{ Behavior of a polyelectrolyte drifting along
a surface with pores in the presence of an applied field. 
(a) The chain tumbles along the surface. (b) It gets
hooked. (c) It stretches and disentangles from a pore. 
(d) Ocassionally after (c), the leading end will thread
itself through a pore. (e) After (d) it is able to
pass to the other side.  }

\label{fig:screen1}
\end{figure}

If the field is
made strong and at an angle, almost parallel to the
surface, then the chain will find it difficult to
pass through a pore. As the angle of the field becomes
parallel and the electric field goes to infinity, the
number of chains that pass through the surface go to zero.
If the chain starts to pass
through a pore, it will occasionally get hooked as shown
in figure \ref{fig:screen1}(a). A chain initially on the
upper side of the screen is shown.   At this point, the
chain stretches, (b), creating a large tension which
yanks the chain out of the pore (c). This was seen in
simulations using a method described in detail below.
This would suggest that the passage of
long molecules would be greatly diminished by this effect.

However upon performing simulations, we also discovered
that there is another mode of motion which
allows long chains to occasionally pass.
Once the chain has stretched as in (c), the leading
end sometimes will pass though a pore (d), and the
chain can thread itself through to the other side (e).
This mode of motion is frequent enough to prevent
this configuration from being an efficient separation
technique. So although the total permeability was
small, the dependence on chain length was still
fairly weak.

We tried using baffles on the lower side of the screen
to prevent this threading mode, but it was not effective.
Once chain is threaded through a pore, it will continue
to migrate through, even if it is running into a wall.

\begin{figure}[tbh]
\begin{center}
\                
\psfig{file=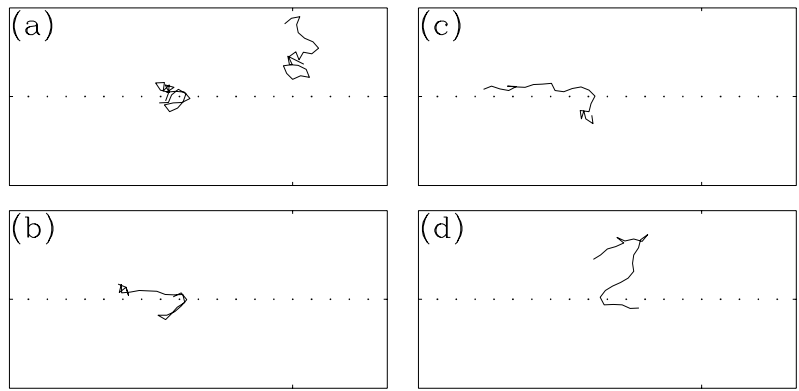,width=5in}
\end{center}
\caption{ Snapshots from the simulation described in
the text of a chain near a porous surface. (a) At the
start of a cycle, the chain is initially away from
the screen, but quickly migrates onto it. (b) The
chain becomes threaded through a pore. (c) It starts
threading and begins to disengage from the surface.
(d) The field is reversed and it migrates away from
the surface.}
\label{fig:snapshots}
\end{figure}

Instead we found that the following sequence  of field 
pulses proved highly effective. Pulses were also crucial
in improving gel electrophoresis\cite{schwartz,chu} for reasons that
are similar to those found here. A set of snapshots
from the simulation are shown in figure \ref{fig:snapshots}. The
 length in monomer units, is $25$.                           
\begin{enumerate}
\item{At the beginning of a cycle the chain is close to the
screen and is in a random coil configuration. This is shown
in  figure \ref{fig:snapshots}(a) labeled ``start".}
\item{Then a field is applied, in these simulations, at a
grade of 2 to 1. It hits the screen as shown in \ref{fig:snapshots}(a)
and then  flattens out and tumbles along the screen.
As in the previous discussion, we see it often becomes hooked
around a pore, see figure \ref{fig:snapshots}(b) and (c).}
\item\label{item:reverse}{The field is suddenly reversed. In these simulations,
for half the duration of the previous stage. The chain
quickly moves of the surface of the screen, \ref{fig:snapshots}(d)}
\item\label{item:relax}{The field is turned off and the chain goes back to
the shape of a random coil}
\end{enumerate}

In the cycle shown in the  figure the chain 
doesn't get through because the field
is reversed before it has a chance to  pass completely through.
Often however it does manage to pass through the screen,
in which case it is unlikely to be able to return to the
other side, since the duration of the
field reversal is for only a fraction of the time in
the  forwards direction.

The point of step \ref{item:reverse}, is to disentangle chains
from the screen and move them away from it. There are more sophisticated
ways to do this which may work better in practice. Instead of only
reversing the field, a large electric could be  first applied parallel
to the surface. This has the effect of disentangling chains
without letting a significant number from the lower side
pass through. At that point the field could be reversed and
this would insure that no chains were entangled with the surface.

The chains need to be a large distance, compared with the
radius of gyration,  from the surface before
they are allowed to relax to random coils. Otherwise 
they can diffuse by Brownian motion to the surface while the coils are 
relaxing. Short and long chains migrate at almost the same rate
in free solution as they act as free draining chains\cite{duke}.
Therefore all the chains will move away from the surface by
about the same distance.
The relaxation time should be that for the longest chains,
or at least it should be set at a relaxation time appropriate
for chains longer than those that we desire to separate.
A simple estimate shows that for the longest chains,
this criterion is met for the design analyzed here.
A chain drifts of order its radius of gyration 
in one relaxation time and the radius of gyration $R_g =  N^{3/5} a$,
$a$ being close to a persistence length and $N$ being the number of persistence
lengths.  Denoting the distance diffused by a chain
$x$, we want to move the chains at a distance away from the screen $d \gg x$.
In one relaxation time a chain moves a distance of order its
radius of gyration, $x = N^{3/5}a $ so  we require
\begin{equation}
d >>  N^{3/5} a ~~.
\label{eq:d}
\end{equation}

To estimate $d$, 
we use the fact that the time-scale for migration
of a chain through the screen is such that a free chain will move
of order its own length\cite{deutsch87}, or several times that. 
Therefore
the reverse field will move a chain of order $N a$.  Therefore 
we need $N a >>  N^{3/5} a $, or
\begin{equation}
N^{2/5} >> 1
\end{equation}
This is well satisfied for chains longer than 100 kilobases.

With a mixture of different chain lengths, the shorter
ones will diffuse more rapidly and have a larger population
at the surface. 
In a time $T$, their density  $\rho (N)$
is proportional to $\exp (-d^2/DT)$ where $D$, 
the diffusion coefficient is proportional  $1/R_g \propto N^{-3/5}$.
Therefore 
\begin{equation}
\rho(N) \propto \exp (-{\rm const.}N^{3/5}) .
\label{eq:exp}
\end{equation}
So even if chains are present at the surface, their numbers decay
very rapidly with chain length, so this effect by itself should
provide excellent resolution.

In addition, if we wish to eliminate this effect, 
a large electric field parallel to the surface can
be maintained during the relaxation stage, \ref{item:relax}.
This inhibits any stray chains near the surface from
passing through it. Then chains substantially shorter
than 100 kilobases should be usable with this apparatus.

\section{Simulation}

\begin{figure}[tbh]
\begin{center}
\                
\psfig{file=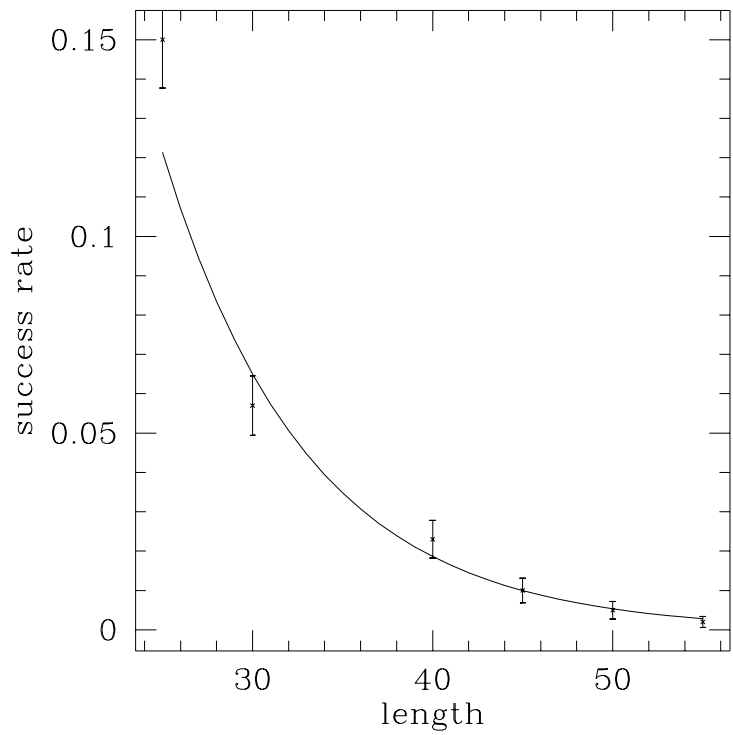,width=4in}
\end{center}
\caption{ The fraction of cycles where the chain completely
passes through the screen. The chain length is given in
units of the number of monomers. The solid line is an
exponential fit to the data.}
\label{fig:permeability}
\end{figure}

Simulation results are shown in figure \ref{fig:permeability}.
The simulation technique used was similar to one used
previously to study gel electrophoresis\cite{deutsch88,deutsch89},
and has had a number of successes predicting, or explaining
the results of experiments\cite{deutsch88,deutsch89,deutschbiref,madden}
on pulsed field gel electrophoresis\cite{schwartz,chu} most 
importantly predicting the motion later seen by fluorescence
microscopy\cite{aldridge,koval,houseal}.

The polymer is represented as a freely hinged chain, with 
the distance between adjacent monomers constant.
An off-lattice Langevin equation was solved for the chain. 
The dynamics were that of a chain in the free draining limit,
and no long range hydrodynamic interactions were included.
Tension between neighboring monomers was introduced to
keep their separation constant. By solving a tridiagonal
matrix equation for each time step, the tensions were
computed, enabling one to update the position of the monomers quite
efficiently.

The screen here is represented by a linear array of pegs.
We specified
$\bf F_i $, the repulsive force between the ith peg and a bead on the
chain. $\bf F_i $ was chosen so as not
to allow crossing of chain through pegs.
Several forms for $ \bf F_i $ have
been tried for gel electrophoresis, all giving similar results. 
The one used in these simulations is the same as was used previously
and has the form
\begin{equation}
{\bf F } ( {\bf r}_i ) ~=~ 
\left\{
\begin{array}{l}
\sigma {\bf \Delta r} 
\left[ {1 \over  \Delta r^2 ~-~ {r_c}^2} - {1 \over 3 {r_c}^2}
\right]^2 \; \mbox{for} \; r_c ~<~ \Delta r ~<~ 2 r_c\\
0\; \mbox{otherwise}
\end{array}\right.
\end{equation}

Here $ \bf \Delta  r $ is the difference between the position
of a bead and the center of a peg.
$ \sigma $ adjusts the strength of the potential, and $ r_c $ represents the hard core radius,
which the chain can never penetrate. For the results described in this paper, $ \sigma ~=~ 0.195 $,
and $ r_c ~=~ 0.2 $, the link lengths equal $0.47$, and the distance
between the pegs equals $1$. This corresponds to having about two
persistence lengths of DNA between pores. With a larger pore separation,
the results described are not expected to change qualitatively.

A random force was added representing thermal noise, and an electric field added
as described above. The magnitude of the parameters chosen is
fairly high for those typically 
used in electrophoresis experiments, fields of about $20V/cm$, however
lower fields should give similar behavior, but take longer to simulate.

In the beginning of a cycle, a chain was initially started in the shape of a 
random walk. The pegs were placed in the x-y plane along the line y=0 with a
spaced by one unit.
Then a downwards electric field in the (-2,-1) direction
was applied for 160 time steps, then it was reversed for 80 time steps. If
all monomers of the chain passed to the other side of the screen,
the cycle was terminated and the passing of the chain through the screen
was recorded.

The results of simulations at different chain
lengths are shown in figure \ref{fig:permeability}.
1000 cycles were performed for each data point.
For long chain lengths the decrease is very sharp. The solid
line is a fit to an exponential. The data appears
to be falling off as an exponential.

Unlike the case of a constant applied field, the use of a pulsed
field appears to enhance the resolution of this device significantly.
There is some average time for the chain to get through
the screen. Such dynamics have been investigated previously
in connection with pulsed field gel electrophoresis\cite{deutsch87}. 
The time scale for crossing the screen should be linear with the
chain length. With a fixed duration on contact with the screen
the probability of it passing through becomes very small. 

A rough estimate of this probability is as follows. In order
for a long chain to pass through, it must initially
be highly stretched so that it can thread its way through quickly.
The probability that a random walk has a radius of gyration
comparable with chain length $N$ decreases exponentially with $N$.

\section{Discussion}
Recently an artificial array of pegs has been 
manufactured out of $\rm SiO_2$ and used in 
studies of electrophoresis\cite{volkmuth,duke}. 
DNA molecules separate well with such arrays. 
It would be interesting to try substituting
a one dimensional array of
pegs as was studied numerically in the last section.

Since such arrays are already
realizable it does not seem too far-fetched to imagine
arrays of pores, instead of pegs, etched in silicon $\rm SiO_2$. With
enough structural support, such arrays would then be usable
on an entire two dimensional surface allowing much more material
to pass through. 

We envisage that in a the complete apparatus will have a large
volume of chains above and below the surface. There
are many possible scenarios for for how the
complete apparatus should work. We discuss below
two ideas that may be useful in obtaining a
realistic working device.

Chains are 
introduced into the upper half and will selectively
migrate to the lower half. Once a chain is in the lower
half, it quickly migrates away from the surface, as the
average field points down. Even if a chain
does reenter the upper half, this will be a sharply decreasing
function of chain length with an even stronger chain
dependence than eqn. (\ref{eq:exp}) and will not adversely effect the
resolving power.

We first discuss a situation where the vertical thickness
of the  upper
half of the device is quite small, of order
of $d$, see eqn. (\ref{eq:d}).

In this case the  composition of chains in the upper half of the device
will change quickly with time.
If enough pulses are applied, all
the chains on the surface will get through and the region
near the surface will be depleted of chains. At some point
we repopulate the chains.
If the upper half is repopulated after $m$ cycles,
by introduction of more solution,
then we can calculate, in steady state, how the
concentration of chains, relative to the upper 
half, depends on chain length. If the probability
that a chain gets through in one cycle is $p$,
then the probability that it will get through
after $m$ cycles is
\begin{equation}
P_m = 1- (1-p)^m
\label{eq:pmcycles}
\end{equation}

\begin{figure}[tbh]
\begin{center}
\                
\psfig{file=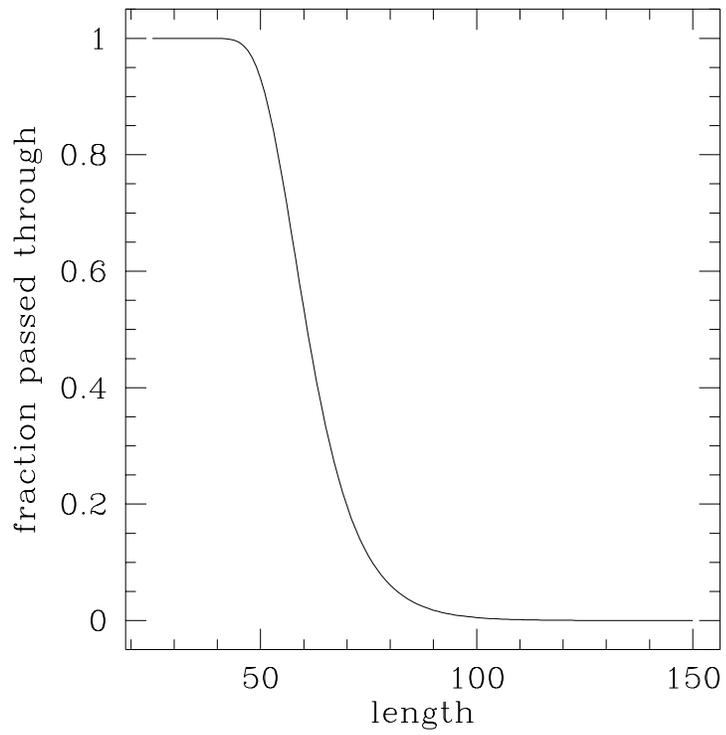,width=4in}
\end{center}
\caption{ The fraction of chains that pass through a screen
after 500 cycles.}
\label{fig:effective}
\end{figure}

To gain an idea of how such an effect enhances
resolution it is necessary to use the exponential
fit of figure \ref{fig:permeability} as simulation
results are not accurate enough at the chain
lengths we now consider. Using this fit for the probability
p as a function of chain length, we have computed
$P_m$ from eqn. (\ref{eq:pmcycles})
for $m= 500$ cycles. This is shown in
figure \ref{fig:effective}.
This has the fortunate effect of making the resolution
even sharper. 

It would be preferable not to have such a shallow
upper half, as the periodic introduction
of more chains complicates the procedure. So consider now the
opposite limit of large vertical depth. In this case, with
an average field pointing down, there is a problem.
Long chains  will pile up at the surface after many pulses.
Therefore it would be better to have an average field
of zero to overcome this difficulty. Note that it is possible to
have an average field of zero and still separate chains.
We refer the reader back to the last paragraph of section
\ref{sec:holes} where we suggested applying a large electric
field almost parallel to the surface to inhibit
possible drift of chains through pores during the
relaxation period. If a large field
that is almost parallel to the surface but tilted slightly
vertically is applied, this will restore a uniform chain
density in the upper half of the device.
During this time there will be negligible
drift of chains in the reverse direction,
as discussed earlier. 

To check that this inhibition takes place, we tried varying
the strength of the vertical field in our
simulation and measuring the
the time it takes for a chain to pass through the
screen in a constant field, for a chain of length
25. When the strength of
the vertical field is decreased so the ratio of
fields is 10 to 1, instead of 2 to 1, the time
it takes to pass through the screen goes up by
about two orders of magnitude. It is so low that it
is very hard to gain adequate statistics to measure it
accurately.

As time
progresses, the depleted region will increase in size
as the square root of time, due to diffusion of chains
from far above the screen. As in the first scenario,
we would like to repopulate chains on the surface.
This could be accomplished by slow stirring, leading
to similar behavior to figure \ref{fig:effective}. 

\section{Screens with Barbs}

The screen studied in the preceding sections does not appear
to be as selective as gel electrophoresis. However we
speculate on how modifications of this basic idea might
allow one to obtain similar resolution by using a more
elaborate surface and a more complicated sequence of
electric fields.

\begin{figure}[tbh]
\begin{center}
\                
\psfig{file=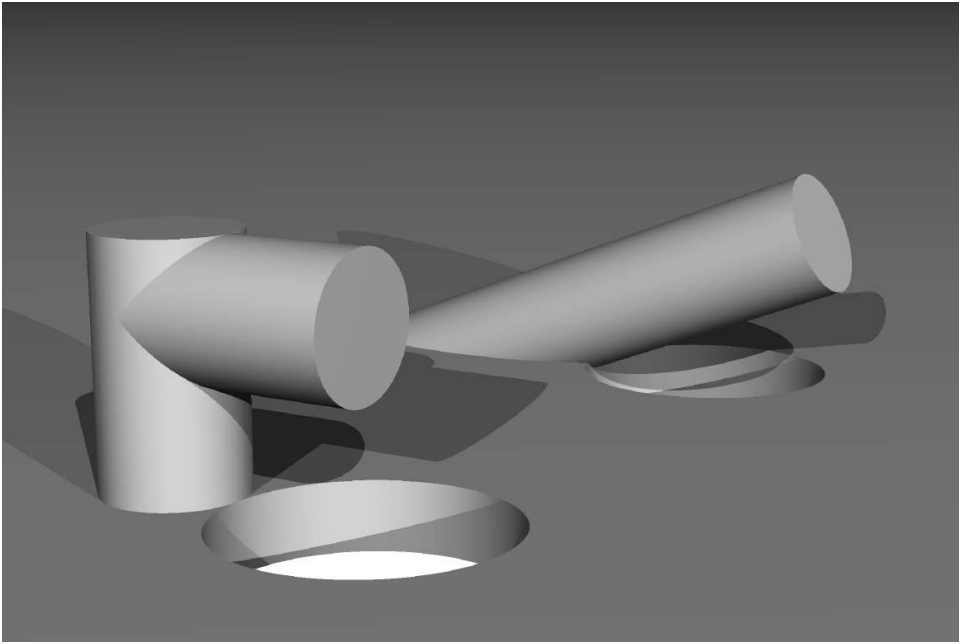,width=5in}
\end{center}
\caption{ Two possible ways of constructing a barb. In the
front, two cylinders at right angles are fabricated. Beneath
them is a pore. In the back, A cylinder is manufactured at
an oblique angle. Beneath it is a pore.}
\label{fig:device}
\end{figure}

Figure \ref{fig:device} shows what is required to make this
device. Besides having holes, cylinders
with overhangs are created next to them as is shown
by the structure in the front. Behind this is an alternative,
cylinders that are tilted towards the horizontal.

The idea behind such a device is that static friction prevents
a chain trapped in a U shape around a barb from moving if
the two arms of the chain have almost the same length. This
effect was predicted theoretically \cite{deutsch87} and
has been seen in fluorescence microscopy experiments\cite{smith}.

\begin{figure}[tbh]
\begin{center}
\                
\psfig{file=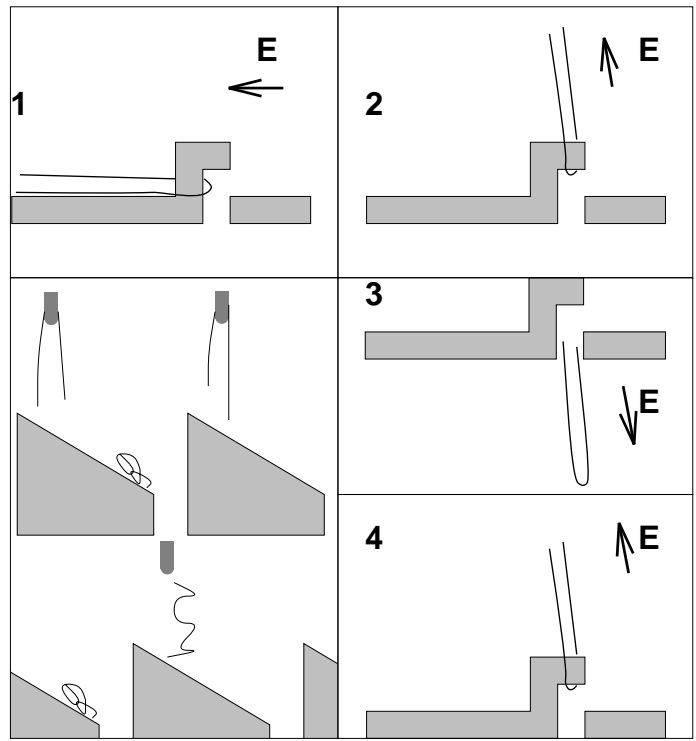,width=4in}
\end{center}
\caption{ {\it Lower left hand diagram:} The device
looking vertically down onto the screen. Chains are
occasionally hooked around barbs but can fall off
landing on the baffle below. It then will curl up
and slide off onto another barb. 1-4 show the sequence
of pulses applied to the device as described in the text.}
\label{fig:screen2}
\end{figure}

A mixture of chains of different lengths, are pushed towards
the surface by an electric field and then the electric field
is directed almost parallel to the surface  so that the chains
migrate. The lower left hand picture in figure \ref{fig:screen2}
shows the surface. Besides the array of barbs, there are
baffles placed underneath them to hasten the curling of chains.
Once a chain has fallen off a barb, it will fall onto a baffle
and curl as it slides down. It will then fall on top of 
another barb where it has a large chance of being hooked by it.

The sequence of fields and motion of the chain during a cycle
is as follows.
\begin{enumerate}
\item{Occasionally the two arms of the chain are close enough
together that the chain is pinned by static friction.
The electric field is continuously rotated until it
is pointing almost vertically up}
\item{It is kept in this configuration for enough time to
allow chains that are not pinned to slide off the barb.
Now only those chains that have their lengths almost equal
are left on the surface. The rest have migrated away from
the surface.}
\item{The electric field is suddenly reversed. The chains
on the surface now enter the hole beneath them. The duration
of this portion of the cycle is crucial in determining
which chains end up on the other side. If the chain is 
too long, the ends of it do not pass all the way through
the hole, otherwise the whole chain passes to the other
side, whereupon the chain ends can diffuse by Brownian motion
laterally}
\item{The field is reversed again. The chain shown is long
enough so that it stays threaded. Therefore it ends up in
the same configuration as in 2. If the chain were shorter.
It would very likely become trapped on the other side.
The important point is that chains longer than a certain
threshold, depending linearly on the duration of step 3,
will almost always end up on the upper surface.}
\end{enumerate}

After stage 4, the field can be set parallel to the surface
towards the right, for a short duration, to allow chains
to come off the barbs. At the point the entire process
repeats itself.

The drawback to this technique is that it requires chains
to become pinned. The number of such chains pinning
per unit time is small, therefore the speed of separation
should be slower than in the previous device that
was analyzed. The advantage is that it is highly  selective
and by adjusting the cycle time as separation progresses,
should be able to extract a very pure samples.

It should be noted that for ring polymers, such a technique
with minor modifications can also be used. In this case it
is  expected to be far more efficient, as a large
fraction of polymers entangled with barbs will be trapped.

\section{Conclusion}

In this paper separation of polymers was considered by using
a permeable surface. The first device considered had
a surface that is an array
of pores of spacing, size and thickness of order several thousand
Angstroms. Improvements in nano-lithography may make such devices
practical. By repeatedly applying a sequence of pulses, it should
differentially separate long for short chains. The critical
chain length where separation occurs should depend linearly on
pulse durations.
One way to test out the ideas proposed in this paper would
be to use a line of pegs similar to previous experiments
on arrays of pegs\cite{volkmuth,duke}. 

The second device would be more complicated to
construct. In addition to the pores, it has an array of barbs
protruding from the surface. Utilizing the fact
that chains  occasionally get pinned by static friction,
a sequence of pulses can be devised to make it possible to very 
precisely separate chains according to their length. This device may be
useful in applications where precise determination of chain
size is needed, but it is not necessary to separate the entire
mixture.

There are still many experimental problems that would need to be addressed 
to make these devices practical. However it might prove worthwhile
investigating these ideas further experimentally.

\section{Acknowledgments}

This work is supported by NSF grant number DMR-9419362
and acknowledgment
is made to the Donors of the Petroleum Research Fund, administered
by the American Chemical Society for partial support of this research.

\end{document}